\documentclass[prd,aps,onecolumn,superscriptaddress,nofootinbib,showpacs,a4paper,10pt, notitlepage]{revtex4-1} 
\usepackage{array}
\usepackage{amsmath}
\usepackage{graphicx}
\usepackage{latexsym}
\usepackage{amsfonts}
\usepackage{pifont}
\usepackage{stmaryrd}
\usepackage{amssymb}
\usepackage{array}
\usepackage{epsfig}
\usepackage[labelformat=default, labelsep=default, justification=centerlast,  font=small]{caption}
\usepackage{verbatim}
\usepackage{multirow}
\usepackage{rotating}
\usepackage{epstopdf}
\usepackage{geometry}
\newcommand{\be}{\begin{equation}}
\newcommand{\ee}[1]{\label{#1} \end{equation}}
\newcommand{\bbe}{\begin{equation*}}
\newcommand{\eee}{\end{equation*}}
\renewcommand{\arraystretch}{1.3}

\def\f12{\frac{1}{2}}
\newcommand{\bra}{\langle}
\newcommand{\ket}{\rangle}

\newcommand{\clock}{\mbox{
\begin{picture}(0,5)
\put(.5,0){\circle{7}}
\put(0.1,0){\line(0,1){2.4}}
\put(0,0){\line(1,0){1.6}}
\end{picture}}}

\newcommand\TspaceBig{\rule{0pt}{3.0ex}}
\newcommand\TspaceBBig{\rule{0pt}{4.0ex}}
\newcommand\Bspace{\rule[-2.0ex]{0pt}{0pt}}
\newcommand\BspaceBig{\rule[-3.0ex]{0pt}{0pt}}

\begin{document}

\title{General relativistic effects in quantum interference of photons}

\author{Magdalena Zych}
\author{Fabio Costa}
\author{Igor Pikovski}
\affiliation{Faculty of Physics, University of Vienna, Boltzmanngasse 5, A-1090 Vienna, Austria}
\author{Timothy C. Ralph}
\affiliation{Department of Physics, University of Queensland, St Lucia, Queensland 4072, Australia}
\author{\v{C}aslav Brukner}
\affiliation{Faculty of Physics, University of Vienna, Boltzmanngasse 5, A-1090 Vienna, Austria}
\affiliation{Institute for Quantum Optics and Quantum Information, Austrian Academy of Sciences, Boltzmanngasse 3, A-1090 Vienna, Austria.}

\date{\today}

\begin{abstract}
Quantum mechanics and general relativity have been extensively and independently confirmed in many experiments. However, the interplay of the two theories has never been tested: all experiments that measured the influence of gravity on quantum systems are consistent with non-relativistic, Newtonian gravity. On the other hand, all tests of general relativity can be described within the framework of classical physics.
Here we discuss a quantum interference experiment with single photons that can probe quantum mechanics in curved space-time. We consider a single photon travelling in superposition along two paths in an interferometer, with each arm experiencing a different gravitational time dilation. If the difference in the time dilations is comparable with the photon's coherence time, the visibility of the quantum interference is predicted to drop, while for shorter time dilations the effect of gravity will result only in a relative phase shift between the two arms. We discuss what aspects of the interplay between quantum mechanics and general relativity are probed in such experiments and analyze the experimental feasibility.
\end{abstract}
\maketitle

\section{Introduction}

Quantum mechanics and Einstein's theory of gravity are the two pillars of modern physics. Since there are conceptual differences between the foundational principles of the two theories, new physics is expected to appear from their interplay at some scale \cite{disconnect}. However, even the regime in which \textit{quantum} systems evolve on \textit{classical}, curved space-time has never been accessed experimentally: all experiments performed so far can be explained either by classical physics in curved background or by quantum mechanics in flat space-time. It is of fundamental interest to verify whether gravitational time dilation applies to single particles in quantum superposition. Likewise, it is important to probe quantum phenomena in regimes where general relativistic effects are present, for example, where time is not a common parameter for different amplitudes of a single quantum system in superposition. In this work we address such a regime and present a scheme which opens a feasible experimental route in this direction.

General relativity predicts that the flow of the time is altered by gravity. This yields several effects that have been independently tested. The gravitational redshift was first observed by Pound and Rebka \cite{ref:Pound1960}, who verified that the frequency of electromagnetic radiation depends on the altitude difference between the emitter and the receiver. In a later experiment, Hafele and Keating \cite{ref:Hafele1972} directly tested both the special and the general relativistic time dilation by comparing actual clocks at different heights moving at different speeds.
Another classical test of general relativity was first proposed and performed by Shapiro \cite{ref:Shapiro1964}: the speed of light, as perceived by a laboratory observer, reduces for electromagnetic waves that travel across regions subject to a gravitational potential \cite{ref:Shapiro1971}.
In these experiments, as well as for all other tests of general relativity, the degrees of freedom relevant for the observation of general relativistic effects can be fully described by the laws of classical physics: the Shapiro effect and the Pound-Rebka experiment can be modelled using classical electrodynamics in curved space-time,
while the Hafele-Keating experiment can be described in terms of clocks measuring time along their classical (localized) trajectories.

The first experiment measuring the effect of gravity on the quantum wavefunction of a single particle was performed by Colella, Overhauser, and Werner (COW) \cite{ref:Colella1975}. In this experiment single neutrons travel in a superposition of two different heights. The different gravitational potential acting on the two paths induces a relative phase to the neutron wavefunction, which is observed from the quantum interference when the two superposed beams are recombined. Today, analogous experiments with atomic fountains are used to perform high-precision measurements of the gravitational acceleration $g$ \cite{ref:Peters1999}. The phase-shifts observed in these interferometric experiments are fully compatible with non-relativistic quantum mechanics in the presence of the Newtonian gravitational potential.
It is perhaps striking that the measurements of gravitationally induced phase-shifts are the only experiments performed to date where both quantum mechanics and gravity are relevant and still they cannot falsify non-metric, Newtonian gravity, which has so thoroughly been disproved for classical systems.


In a recent work \cite{ref:Zych2011}, an experiment was proposed that allows probing general relativistic time dilation in conjunction with the quantum complementarity principle (which states that quantum interference is lost once which-way information becomes available).
As in the COW experiment, one considers  interference of matter waves. The novelty of the scheme is to use additionally some controllable time-evolving degree of freedom of the mater-wave as a clock \cite{ref:Zych2011, ref:Sinha2011}. The system is brought in a coherent superposition of two locations at different gravitational potentials (e.g. at different heights above the earth). As a result of the general relativistic time dilation the rate of the time evolution of the clock will be different for each of the amplitudes of the superposition, depending on its location. Thus, according to quantum complementarity the coherence of such a spatial superposition will drop to the extent to which the information on the position becomes accessible from the state of the clock. In this setup one can therefore probe the general relativistic time dilation by observing the reduction of the visibility of quantum interference, in addition to the observation of the phase shift. The measurement of such a gravitationally induced decoherence (and revivals - for a periodic clock)  would be the first confirmation of a genuinely general relativistic effect in quantum mechanics.

Here we discuss a quantum-optics variation of this proposal where the function of the ``clock'' is taken by the position of a single photon along the interferometer's arm. Because of general relativistic time dilation, the arrival time of the photon should depend on the altitude of its trajectory above the earth (consistently with the Shapiro delay). Thus, in an experiment with a single photon travelling in a superposition along two paths of an interferometer, each located at a different height above the earth, a reduction of the interferometric visibility is expected for relative time dilations larger than the photon's coherence time.
This again can be understood in terms of quantum complementarity: if the which-path information of the photon can be read from its time of arrival, interference is lost.

The proposed effect depends simultaneously on general relativity and on quantum mechanics and therefore goes beyond previous experiments that probed the two theories independently. The experiment can distinguish between a semi-classical and a quantum extension of the mass-energy equivalence. To illustrate the differences between these two we present a toy model (Appendix \ref{app_b}) which incorporates such a semi-classical version of the mass-energy equivalence. The model is different than quantum mechanics in curved space-time, yet still in agreement with all current observations of either classical systems in curved space-time or quantum systems in Newtonian gravity.



The paper is organized as follows: in Sec.\ \ref{sec:masssive_clock} we review the quantum interferometry of clocks aimed at testing the general relativistic time dilation; in Sec.\ \ref{sec:photons} we derive the predicted outcomes for a photonic version of such an experiment;  in Sec.\ \ref{sec:exp_param} the feasibility of the scheme is discussed; in Sec.\ \ref{sec:discussion} we analyze which aspects of general relativity and quantum mechanics are tested in our proposal and in other experiments and address possible related interpretational issues; we conclude in Sec.\ \ref{sec:conclusion}.

\section{Quantum interference of clocks}
\label{sec:masssive_clock}
\subsection{The setup}
In this section we briefly review the proposal given in Ref.\ \cite{ref:Zych2011}.
%
\begin{figure}[h!]
\vspace{0pt}
\begin{center}
\includegraphics[width=9.0cm]{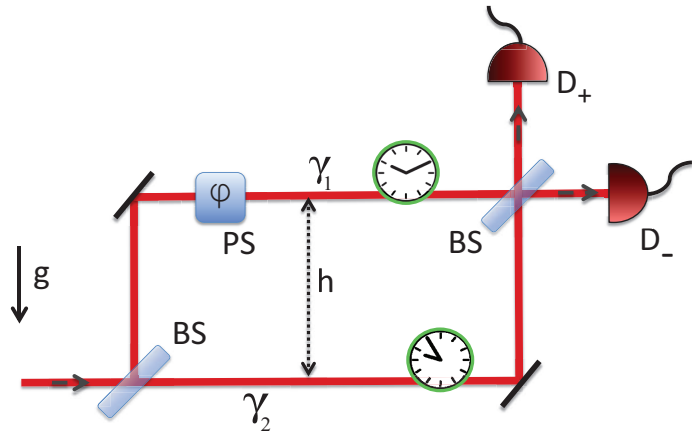}
\end{center}
\caption{A Mach-Zehnder interferometer to measure the effect of general relativistic time dilation on a wavefunction of a single quantum system. The setup is placed in a homogeneous gravitational field ($g$) and it consists of two beam splitters (BS), two detectors $D_\pm$ and the phase shifter (PS) which gives a controllable phase difference $ \varphi$ between the two trajectories $\gamma_1$ and $\gamma_2$. 
The separation between the paths in the direction of the field is $ h$. If the quantum particle that travels in superposition keeps track of time along its path, the visibility of the interference will be reduced since the which-way information becomes accessible due to the time dilation between the paths.
\hspace*{\fill}}
\label{machzehnder}
\end{figure}
A Mach-Zehnder interferometric setup for matter-waves is placed in the earth's gravitational field (see Fig.\ \ref{machzehnder}), and the massive particle is put in a coherent superposition of travelling along the two paths $\gamma_1$ and $\gamma_2$. The particle has additional internal degrees of freedom that can serve as a clock (e.g. a spin precessing in a magnetic field), which evolve into states $|\tau_i\ket$, $i=1,2$ on the respective trajectories $\gamma_i$. Additionally, the particle acquires a trajectory dependent phase $\Phi_i$ and a controllable relative phase shift $\varphi$. The external modes associated with the paths are denoted by $|r_i\ket$. 
%
%
After  the superposition is recombined, the particle can be registered in one of the two detectors  $D_\pm$. The state of the particle inside the interferometer reads
\begin{equation}
\label{intermed_tau}
|\Psi\ket=\frac{1}{\sqrt{2}}\left(ie^{-i\Phi_1}|r_1\ket |\tau_1\ket +e^{-i\Phi_2+i\varphi}|r_2\ket |\tau_2\ket \right).
\end{equation}
According to general relativity, time is not absolute but flows at different rates depending on the geometry of space-time. This means that the clock will evolve into different quantum states, depending on the path taken. The state \eqref{intermed_tau} is then entangled and, according to quantum mechanics, interference in the path degrees of freedom should correspondingly be washed away. The reason is that one could measure the clock  degrees of freedom and in that way access the which-path information. Indeed, tracing out  the clock states in Eq.\ \eqref{intermed_tau} gives the detection probabilities $P_\pm$ associated with the two detectors $D_\pm$:
\begin{equation}
\label{probab_intro_tau}
P_{\pm}=\frac{1}{2}\pm \frac{1}{2}|\bra\tau_1|\tau_2\ket|\cos\left(\Delta\Phi+\alpha + \varphi \right),
\end{equation}
where $\bra\tau_1|\tau_2\ket = |\bra\tau_1|\tau_2\ket|e^{i\alpha}$ and $\Delta\Phi:=\Phi_1-\Phi_2$. When the ancillary phase shift $\varphi$ is varied, the probabilities $P_\pm$ oscillate with the amplitude $\mathcal V$, called the visibility of the interference pattern. Formally $\mathcal V:=\frac{Max_{\varphi} P_\pm - Min_{\varphi} P_\pm}{Max_{\varphi} P_\pm + Min_{\varphi} P_\pm}$  and for the case of Eq.\ \eqref{probab_intro_tau} it reads
\begin{equation}
\label{visib_tau}
\mathcal V=|\bra\tau_1|\tau_2\ket|.
\end{equation}
The visibility of the interference pattern depends on the distinguishability of the clock states that followed different interferometric paths. If we quantify the amount of the which-way information $\mathcal D$ (distinguishability of the trajectories) as the probability to correctly guess which path was taken in the two-way interferometer \cite{englert}, we obtain $\mathcal D = \sqrt{1-|\bra\tau_1|\tau_2\ket|^2} $. The complementarity principle can then be phrased in the form of the well-known duality  relation $\mathcal V^2+\mathcal D^2=1$, see Refs.\ \cite{englert, green_yasin, wooters}.

In Newtonian gravity the expected visibility is (ideally) always maximal ($\mathcal V = 1$) because the rate of the time evolution of the clock is the same along both paths, i.e.\ $|\tau_1\ket=|\tau_2\ket$. In this limit the relative phase shift between the two amplitudes is the only effect of gravity and the internal clock is irrelevant. The phase $\Phi_i$ is proportional to the action along the corresponding (semiclassical) trajectory $\gamma_i$ on which the particle moves. In the presence of the position dependent potential $V(x)$ this phase becomes trajectory dependent: $\Phi_i = m\int_{\gamma_i} V(x(t))dt$ for a particle of mass $m$ (assuming that contributions from the kinetic energy cancel out). Therefore, even in a homogeneous field the particle acquires a trajectory dependent phase although the force acting on it is the same at all points - the phase arises only due to the potential. For a homogeneous electric field such a relative phase is sometimes referred to as the electric Aharonov-Bohm effect \cite{aharonov}. The case of Newtonian gravity is directly analogous - the role of the particle's electric charge and of the Coulomb potential are taken by the particle's mass and the Newtonian gravitational potential, respectively \cite{morgan}.

Note that Newtonian gravity is a limiting case of Einstein's theory of gravity. Any Newtonian effect can therefore be equivalently described in the general relativistic framework. Within this framework, the action of a particle only subject to gravity is proportional to the proper time elapsed along the particle's trajectory: $S_i=-mc^2 \int_{\gamma_i}d\tau$. The low-energy limit of this expression is the action of a non-relativistic massive particle subject to the Newtonian gravitational potential. Thus, even the Newtonian interferometric phase shift can be equivalently expressed as $\Delta\Phi = \frac{m c^2}{\hbar}\Delta \tau$ where $\Delta \tau$ is the proper time difference between the trajectories. Some authors interpret  therefore the gravitational phase shift as the effect of gravitational time dilation \cite{ref:Mueller2010}. Even though such an interpretation is consistent with general relativity, it cannot distinguish Newtonian (i.e. non-metric) gravity from genuine general relativistic effects as long as only the Newtonian limit of the phase shift, $\Delta\Phi = m g h t/\hbar$, is measured (as in the experiments performed so far). Similarly, any experiment that probes the acceleration of free fall, like Galilei's leaning tower experiment, cannot distinguish the effects of Newtonian gravity from general relativistic effects, although it can be interpreted as arising from gravitational time dilation when formulated within the general relativistic framework. An interpretation of these experiments as measuring the time dilation can only be sustained if the very existence of the time dilation is pre-assumed, i.e.\ by pre-assuming that the observed Newtonian effects are the low-energy limit of general relativistic effects. 

Additionally, to sustain the general relativistic viewpoint it has been argued that, differently from classical free fall experiments, a massive quantum particle constitutes a ``clock'' ticking at the Compton frequency, which causes the interference. The notion of such a ``Compton clock'', however, is purely formal: no measurement on the particle within the interferometer can reveal the time of such a clock\footnote{The claim that the measurement of the phase shift itself, being \textit{proportional} to the Compton frequency, is a measurement of the latter is untenable: any measurement of the mass $m$ is likewise proportional to the Compton frequency.}. The phase shift observed is always just a relative phase between the two trajectories. In order to measure the frequency of the ``Compton clock'', it would be necessary to measure the number of oscillations per unit time of the global phase of the wavefunction, in contradiction to the basic tenets of quantum mechanics.
Moreover, any relative phase shift can be explained as arising due to an effective (possibly even non-Newtonian) gravitational potential within a non-metric theory, which does not predict the effect of time dilation.
To conclude, such measurements can be used to test possible deviations from general relativity in the Newtonian limit \cite{ref:Mueller2010,  Haensch:2004} but
not to distinguish general relativity from Newtonian theory.
When performed with higher precision, phase shift measurements alone can also reveal possible deviations from the Newtonian gravitational potential \cite{dimopoulos, chu:2008}, but will always be compatible with an absolute time (flat space-time).
In contrast, the drop in the interferometric visibility (only observable when using \textit{physical} clocks),  cannot be explained without the general relativistic notion of proper time. 

\subsection{Interferometric visibility and which way-information}
\label{sec:visibility}
In order to explain the reduction in the interferometric visibility both quantum complementarity and the general relativistic time dilation are necessary.  
Here the which-way information becomes  available only due to the gravitational time dilation. The time evolution of the upper amplitude of the clock precedes the lower one by $\Delta \tau$ - the time dilation between the paths of the interferometer. The distinguishability, however, depends on how $\Delta \tau$ compares with the precision $t_\perp$ of the clock - i.e. the time that the quantum system needs to evolve between two distinguishable states. The expected visibility will thus depend only on the ratio of these two parameters and Eq.\eqref{visib_tau} will generally take the form:
%
\begin{equation}
\label{visib_general}
\mathcal V=F_{\clock}\left(\frac{\Delta \tau}{t_{\perp}}\right),
\end{equation}
where the function $F_{\clock} \,$ has the following properties: $F_{\clock\;}: R_{+} \rightarrow [0,1]$ such that  $F_{\clock\;}(0)=1$ and $F_{\clock\;}(1)<<1$. Note that this conclusion is not limited to any particular physical implementation of the clock. The fact that the time dilation necessary to cause a loss of quantum interference only depends on the precision of the clock and not on its realization can be seen as a consequence of the universality of general relativistic time dilation: rates of all clocks are equally affected by gravity. Only the explicit form of the function $F_{\clock\;}$ depends on the specific realization of the clock. For example, a clock  with a finite dimensional Hilbert space has a periodic time evolution and thus one expects periodic modulations of the visibility with increasing time dilation between the two arms of the interferometer. For a clock implemented in a two-level quantum system the visibility reads \cite{ref:Zych2011}
\begin{equation}
\label{visib2}
\mathcal V=\left| \cos \left( \frac{\Delta \tau}{t_{\perp}}\frac{\pi}{2} \right) \right|.
\end{equation}
For such a  periodic clock, consecutive states at multiples of $t_\perp$ are mutually orthogonal (and in this case the parameter $t_\perp$ is known as the orthogonalization time), so the full loss of the visibility is expected for $\Delta \tau = t_\perp$. The period of a two-state clock is twice $t_\perp$, which explains the full revival of the visibility for $\Delta \tau = 2t_\perp$. Such a clock may be realized in hyperfine energy levels of an atom or with the spin of a neutron.

\section{Testing general relativistic effects with single photons}
\label{sec:photons}
We now consider a variation of the interferometric experiment that can be performed with single photons. In this case, the clock is implemented in the position degree of freedom of the photon (which has an infinite dimensional Hilbert space).
We first consider the light-like geodesics within an interferometric setup as in Fig.\ \ref{machzehnder} and then consider a single photon travelling in superposition along these geodesics. The interferometer is placed, vertically oriented, on the surface of the earth and the space-time is modeled by the Schwarzschild metric \cite{BIR82}.
%
%
For a small size of the interferometer as compared to its radial coordinate $r$, all points on each of the horizontal paths are approximately at the same radial distance and the vertical distance $h$ can be adjusted whilst keeping the horizontal path length $l$ constant. We therefore restrict our attention to horizontal propagation in the $x$ direction and describe the motion in an approximately flat  $2D$ metric
\begin{eqnarray}
ds^2& =& d\tau_r^2-\frac{1}{c^2}dx_r^2 \;,
\label{M2}
\end{eqnarray}
where
\begin{equation}
d\tau_r^2 = \left(1+\frac{2V(r)}{c^2}\right)dt^2\;,\;\;\;d x_r^2 =  r^2 d\theta^2\;.
\label{SF}
\end{equation}
$V(r) = -\frac{GM}{r}$ is the gravitational potential, $G$ is the universal gravitational constant and $M$ is the earth's mass.  The coordinates are defined as follows: $dt$ is an infinitesimal time interval as measured by an observer far from the earth, $\theta$ is a polar coordinate that remains the same for the observer on earth and the far away observer, $\tau_r$ and $x_r$ are the local time and the local cartesian coordinate, respectively, for an observer at radial coordinate $r$. We first focus on the coordinate time $t_r$ of the photon's path along a horizontal trajectory at the radial distance $r$. Since the trajectory in this case is light-like, i.e. $ds^2=0$,  Eq.\ \eqref{M2} results in $t_r = \frac{1}{c} \int_0^{l} dx_r \left(1+\frac{2V(r)}{c^2}\right)^{-1/2}$.
%
%
By definition of the metric, the locally measured spatial intervals are independent of $r$ and $\int_{0}^{l} dx_r = l$, so that
\begin{equation}
\label{coord_time}
t_r =  \frac{l}{c \sqrt{1+\frac{2V(r)}{c^2}}}.
\end{equation}
According to Eq.\ \eqref{coord_time}, the time of the photon's flight along the horizontal path can be seen as a clock, which is subject to the gravitational time dilation as predicted by general relativity (which is the Shapiro effect). Depending on the radial distance $r$ of the path from the earth, the photon will arrive at the second beam splitter at different coordinate times. From the symmetry of the setup,  the coordinate time of the photon's flight in the radial direction is the same for both trajectories $\gamma_i$. The total difference in photon arrival times as measured by the far away observer is therefore $t_r-t_{r+h}$. For the local observer at the upper path (at the radial distance $r+ h$) this time dilation is given by
\begin{equation}
\label{time_diff}
\Delta \tau  =  \sqrt{1+\frac{2V(r+h)}{c^2}}\left(t_r - t_{r+h}\right) \approx \frac{lgh}{c^3}\;,
\end{equation}
with $g=\frac{GM}{r^2}$.  The approximation in Eq.\ \eqref{time_diff} is valid for a small size of the interferometer, i.e. $h<<r$, and when second order terms in the potential are neglected.

To probe quantum mechanics in curved space-time, we consider the above effect on single photons in superposition. The state of a photon moving in the $+x$ direction at the radial distance $r$ in the space-time given by the metric \eqref{M2} can be written as
\begin{eqnarray}
|1 \rangle_f = a^\dagger_f |0 \rangle =\int d\nu f(\nu) e^{i\frac{\nu}{c}(x_{r} - c \tau_{r})} a^{\dagger}_{\nu} |0 \rangle,
\label{sig}
\end{eqnarray}
where $f(\nu)$ is the mode function with $\nu$ being the angular frequency defined with respect to a local observer at radial distance $r$ and $a^{\dagger}_{\nu}$ being a single frequency bosonic creation operator. According to Eq.\ \eqref{time_diff} the wave packet moves along each of the trajectories with a different group velocity with respect to a fixed observer. Therefore, the superposition states reaching the two detectors $D_\pm$ take the form
 $|1 \rangle_{f\pm} \propto \int d\nu f (\nu)\left( e^{i\frac{\nu}{c}(x_{r} - c \tau_{r})} \pm e^{i\frac{\nu}{c}(x_{r} - c (\tau_{r}+\Delta \tau))} \right) a^{\dagger}_{\nu} |0 \rangle$, with $\Delta \tau$ as given by Eq.\ \eqref{time_diff}. The probabilities of detecting the particle in $D_\pm$ read (see Appendix \ref{app_a} for a supplementary calculation)
\begin{eqnarray}
P_{f\pm}  = \frac{1}{2}\left(1 \pm \int d\nu |f(\nu)|^2 \cos(\nu\Delta \tau)\right).
\label{photon_probab}
\end{eqnarray}

The mode function is normalized such that $\int d\nu |f(\nu)|^2 = 1$. We can identify two limits of this expression, which in turn correspond to tests of two different physical effects:

(i) For $\Delta\tau$ small with respect to the frequency-width of the mode function $f(\nu)$ the cosine term will be approximately constant over the relevant range of $\nu$ and so can be taken outside the integral giving a phase shift term $\cos( \nu_0 \Delta\tau)$, (where $\nu_0$ is central frequency of the mode function) and the visibility remains maximal.
Note that  such a phase shift can be explained as arising from the coupling of the average energy of the photon to the Newtonian gravitational potential in the Euclidean space-time, where the time is absolute and no time dilation occurs --- analogously to the case of a massive particle. The gravitational phase shift for a photon is also of the same form as the phase shift for a massive particle, with the photon's energy (divided by $c^2$) substituting the particle's mass. It thus constitutes an interesting test of the mass-energy equivalence and the resulting coupling to gravity, one of the conceptual pillars of general relativity. Although the measurement of a gravitational phase shift cannot be considered as a test of the time dilation, its observation with a single photon differs from a corresponding test for a massive particle because in the Newtonian limit of gravity no effect on a massless system would be expected (see Table \ref{table:summary}). In this sense in such an experiment both quantum and general relativistic effects could simultaneously be tested. Furthermore this proposal is particularly interesting as it seems feasible with current technology\footnote{A similar experiment, where the gravitationally induced phase shift of classical light was considered, was first proposed in 1983 \cite{tanaka}, but has not been performed yet.}.

(ii) In the other limit, when the time dilation dominates over the pulse width, the oscillatory cosine term averages out the whole integrand, resulting in no interference. More precisely, the amplitude of the phase shift term is given by the overlap between the modes associated with the two paths.  For $\Delta \tau$ much larger than the coherence time of the photon wave packet (the pulse width) the interferometric visibility is lost as the two modes arrive at the second beam splitter at distinctly different times.
Note that the slow-down of light is a direct consequence of the gravitational time dilation, and is absent in the non-metric Newtonian theory (see Sec.\ \ref{sec:discussion} and Appendix \ref{app_b} for further discussion). The drop in the interferometric visibility is therefore probing a regime beyond the Newtonian limit. In this parameter regime the experiment directly tests the gravitational time dilation for superpositions of single photons.

The two limits can be clearly seen when considering the specific case of a gaussian wave packet
$f_{\sigma}^{\nu_0}(\nu)=(\frac{\sigma}{\pi})^{1/4}\exp[{-\frac{\sigma}{2}(\nu-\nu_0)^2}]$:
\begin{eqnarray}
P_{f_\sigma^{\nu_0}\pm}  = \frac{1}{2}\left(1 \pm e^{-(\Delta \tau/2\sqrt{\sigma})^2} \cos(\nu_0\Delta \tau)\right).
\label{photon_probab_gauss}
\end{eqnarray}
The width $\sigma$ of the  gaussian mode function gives the precision $t_\perp$ of  this clock. Defining distinguishable gaussian wave packets as such that their overlap is not larger than $e^{-1}$, the precision of the clock is given by  $t_\perp =2\sqrt\sigma$ and the visibility of the interference in Eq.\  \eqref{photon_probab_gauss} reads
\begin{eqnarray}
\mathcal{V} = e^{-\left(\frac{\Delta \tau}{t_{\perp}}\right)^2}.
\label{visibility_gaussian}
\end{eqnarray}

The clock implemented in the position degree of freedom of a photon is thus analogous to the clock implemented in an internal degree of freedom of a massive particle \cite{ref:Zych2011}: the interferometric visibility depends only on the ratio of the time dilation to the precision of  the used clock\footnote{In general, demanding that the two wave packets are distinguishable when the absolute value of the amplitude between them is $1/n$ yields $\mathcal V_n = n^{-(\Delta \tau/t_{n\perp})^2}$ where $t_{n\perp} = 2 \log(n) \sqrt{\sigma}$ is the corresponding clock's precision.} and the phase shift depends on the average energy of the clock, which is the mean frequency for the case of a photon.

\section{Quantitative predictions}
\label{sec:exp_param}
For a single photon in superposition, a full loss of the visibility can be obtained in a sufficiently large interferometer, such that $\Delta \tau > t_\perp$. Assuming the gravitational field to be homogeneous the corresponding area of the interferometer obtained from the Eq.\ \eqref{time_diff} reads
%
\begin{equation}\label{area}
A_{\perp} \approx t_\perp \frac{c^3}{g}.
\end{equation}
For a clock of the same frequency but moving at a subluminal velocity $v$, the corresponding area of the interferometer necessary to observe the loss of the interferometric visibility is by $v/c$  smaller than the area needed for the clock moving at velocity $c$. Thus, the size of the interferometer necessary for realizing the proposed experiment with the clock implemented in a photon is orders of magnitude larger than the setup needed for a massive clock. However, such an implementation can still be a feasible route for the observation of time dilation in a quantum system as an implementation with photons may provide some advantages. For example, the control over the exact length and separation between the interferometric paths would be easier with a photon clock since it could be confined in an optical fiber. Moreover, experiments were already performed confirming the high-fidelity transmission of polarization-encoded qubits from entangled photon pair sources over 100 km in a fiber \cite{ref:Huebel}. For the preparation of a clock with a fixed precision one needs to control only the width of the photon's wave packet. The preparation of light pulses with duration on the order of 10 femtoseconds has already been achieved \cite{ref:Mosley} and recent rapid progress in the preparation of  attosecond optical pulses may provide a feasible route to such high-precision clocks \cite{ref:Sansone, ref:Gallmann}. Finally, using  slow light materials (e.g.\ with a suitable frequency dependent  index of refraction) may allow reducing the necessary size of the interferometer.

\captionsetup[table]{labelformat=default, labelsep=default, justification=centerlast, font=small}
\begin{table}[h!]
\caption{The size of the interferometer necessary to observe a loss of visibility due to the gravitational time dilation  and to observe the gravitational phase shift. For the realization of the clock in the position degree of freedom of a photon the precision is given by the coherence time of the photon's wave packet. For the clock implemented in the internal degree of freedom of a massive particle - by the orthogonalization time (the shortest time after which the final state becomes orthogonal to the initial one). In both cases only the total energy of the system is relevant for the observation of the gravitational phase shift.
\hspace*{\fill}}
\centering
\begin{tabular}{|c| c c c |c| }
\hline
 \multirow{3}{*}{{\bf time dilation}}     & \textbf{system}       &   \textbf{ clock  }       & $\mathbf{t_\perp}$ {\bf[s]} & $\mathbf{A_{\perp}}$ {\bf[km$^2$]} \\
                                                                &  photon                   & position                        & $10^{-15}$                           & $10^{3}$                \\
                                                                &  atom                      & hyperfine states          & $10^{-15}$                            & $10^{-7}$               \\    \hline \hline
\multirow{2}{*}{{\bf phase shift}}              & \multicolumn{3}{c|}{ {\bf photon frequency} \bf[Hz]}              & $\mathbf{A_{ps}}$ {\bf[km$^2$]} \\
                                                                &              \multicolumn{3}{c|}{     $ 10^{15}      $ }                                &    $   10^{-3}  $           \\ \hline

\end{tabular}
\label{table:expparamters}
\end{table}

\begin{figure}[ht]
\vspace{0pt}
\begin{center}
\includegraphics[width=8.5cm]{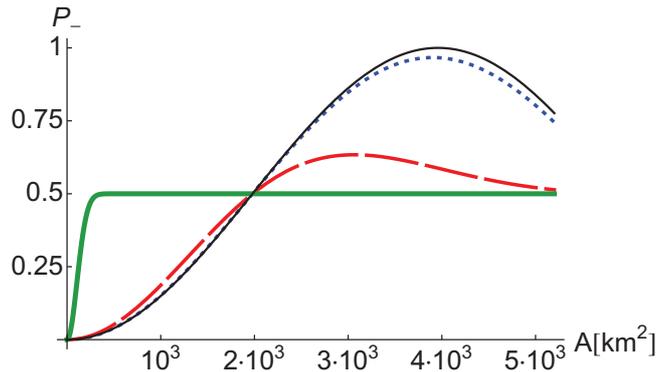}
\end{center}
\vspace{0pt}
\caption{
Probing the general relativistic slowdown of light in the proposed single photon interference experiments. The probabilities $P_{f_\sigma^{\nu_0}-} $ (Eq.\ \eqref{photon_probab_gauss}) to find the photon in the detector $D_-$ are shown as a function of  the area $A$ [km$^2$] of the interferometer in the vertical plane. The single-photon pulse has mean angular frequency $\nu_0 \approx 4\times10^{15}$~Hz. The four graphs correspond to different widths $\sigma$ of the wave packet and thus to different coherence times $t_\perp = 2\sqrt{\sigma}$:  $t_\perp =0.05$~fs (thick green line),  $t_\perp =1$~fs (red dashed line), $t_\perp =5$~fs (blue dotted line) and $t_\perp =\infty$~fs (black line), which is a limiting case of an infinitely long pulse (and thus with the clock effectively switched off). Observing a gravitational phase shift of a photon is within reach of current experimental capabilities.
\hspace*{\fill}}
\label{fig:plot}
\end{figure}

The experimental parameters necessary to measure the gravitational phase shift for single photons are much less demanding than those required for full orthogonalization (see Table \ref{table:expparamters} for a  comparison of the experimental parameters necessary to observe those effects and Fig.\ \ref{fig:plot} for a plot of the detection probability $P_-$, Eq.\ \eqref{photon_probab_gauss}, as a function of the interferometer's area). For a photon with a central frequency $\nu_0$, the gravitational phase shift in the considered interferometer is $\Delta \Phi \approx \nu_0 \frac{lgh}{c^3}$. Taking the angular frequency $\nu_0 = 10^{15}$~Hz, and the time dilation as used above ($\Delta \tau = 10^{-15}$~s) we get the phase shift of order of one radian. Since phase shifts that are at least six orders of magnitude smaller can be experimentally resolved \cite{matsuda2009}, the area enclosed by the interferometer that would suffice to result in the measurable phase shift effect is correspondingly smaller: $A_{ps} = 10^3$~m$^2$. Such a size is feasible with current technology. One could also introduce some improvements to the proposed realization of the experiment. For example, by using multi-photon states or different interference schemes \cite{pryde2011, hom1986} not only would it be possible to enhance the precision of the measured phase but, by observing an interference with no classical analogue, the interpretation of the experiment as testing the mass-energy equivalence in conjunction with quantum mechanics would be unambiguous.


%

\section{Interpretation of photon experiments}
\label{sec:discussion}

The aim of the proposed experiment is to access a physical regime where the effects stemming from both quantum mechanics and general relativity (GR) cannot be neglected.
To this end,  experimental results should simultaneously falsify a model of quantum fields or particles evolving in a Newtonian potential \textit{and} a model of classical fields or particles in curved space-time. Below we  analyze in which sense and how this can be achieved.  Finally, we show how with a modified experimental setup more general local realistic models of light in curved space-time can be ruled out.

\subsection{General relativistic and quantum aspects of the proposal}
In the absence of relativistic effects, clocks run at the same rate independently of their position with respect to a gravitational field.  In the presently proposed experiment, the clock is implemented in the position of the photon in the direction parallel to the earth's surface. Since the gravitational field is constant in that direction, Newtonian gravity would not provide any measurable effect on the time evolution of the clock, i.e. on the displacement of the photon. A difference in the arrival time for the two superposed wave packets of the photon is inconsistent with such a non-relativistic model and can therefore be considered as a genuine general-relativistic effect. Since in our current proposal the difference in the arrival time is measured by the drop in the interferometric visibility, the experiment needs to be able to discriminate between this effect and other sources of decoherence, which could also lead to the loss of interference. This can be done by rotating the interferometer from a vertical to a horizontal position, which should lead to a corresponding recovery of the interference; a correlation between the rotation angle and the visibility would imply that gravity is responsible for the effect. Another possibility is to use wave packets with a modulated profile in the time domain (e.g. with a double or multiple peak): by increasing the time delay the visibility should be partially recovered, as the two superposed wave packets start overlapping again. Since a revival of visibility cannot be attributed to decoherence, this possibility would be ruled out by the experiment.

As discussed above, the effect of gravity can also be detected by observing a shift in the relative phase between the superposed wave packets travelling along the two arms. 
The observation of the phase shift alone could not be directly interpreted as a test of the time dilation, since it can be understood in terms of a coupling between photons and a potential in a flat space-time, which would yield the same phase shift \textit{without} the time dilation. The formal analogy of a massive particle in the Newtonian limit of gravity with a charged particle in a Coulomb electrostatic potential (where notions of proper time or space-time curvature unquestionably never enter) suggests that only those gravitational effects that have no electrostatic analogues can be seen as genuinely general relativistic. The gravitational  phase shift effect does have an electrostatic counterpart (experimentally verified with electrons in a Coulomb potential, see e.g.\ \cite{electrons}) and thus its measurements (such as in Refs.\ \cite{ref:Colella1975, ref:Peters1999}) do not contradict non-metric, Newtonian gravity and cannot be seen as tests of general relativity. The interaction of \textit{photons} with the Newton potential, however, requires an additional ingredient: it is necessary to assign an effective gravitational mass to each photon, whereas massless particles or electromagnetic waves do not interact with gravity in non-relativistic mechanics. 
Such an interaction is a direct consequence of the mass-energy equivalence. More precisely, the interaction necessary to obtain the gravitational phase shift follows from postulating the equivalence between the system's mass and its average total energy, $mc^2\rightarrow mc^2+ \bra \hat H_s\ket$, where $\hat H_s$ is the non-gravitational Hamiltonian of the system. This would yield an interaction of the form
\begin{equation}
\hat H_{phase}=  \bra \hat H_s \ket \frac{V(\hat r)}{c^2},
\label{H_PS}
\end{equation}
which reduces to the Newtonian gravitational potential energy, $mV(\hat r)$, in the non-relativistic limit. In this sense, observation of a gravitationally induced phase shift for a photon ($m=0$) can already be considered as a test of general relativity. In particular, it probes a \textit{semi-classical} version of the mass-energy equivalence, which endows photons with an effective gravitational mass \textit{parameter}\footnote{Note that the Pound-Rebka experiment \cite{ref:Pound1960} probed the same aspect of GR: the redshift can be explained as each photon having a gravitational potential energy of $g h E /c^2$, with $E=\hbar \nu$. The redshift is then a consequence of the (semi-classical) mass-energy equivalence and energy conservation. However, the quantum nature of the photons was not probed directly in the Pound-Rebka experiment, which is thus still consistent with a classical description of light (more precisely, the experiment is consistent with classical waves on a curved background \textit{or} with classical particles in a Newtonian potential, see Table \ref{table:summary}). Measuring the gravitational phase shift in single photon interference would allow probing this aspect of GR in a quantum regime. In simple terms, it would provide a quantum extension of the Pound-Rebka experiment in the same way as the COW experiment was a quantum extension of Galilei's free fall experiments.}: $\;\frac{\bra \hat H_s\ket}{c^2}=\frac{\hbar \nu}{c^2}$.  
The interaction \eqref{H_PS} results in a relative phase between the amplitudes of a wavefunction at different gravitational potentials, but it predicts no drop in the interferometric visibility (see Appendix \ref{app_b}). The latter effect is only present if the total hamiltonian includes a direct coupling between the energy operator and the potential, a term of  the form
\begin{equation}
\hat H_{vis}= \hat H_s\frac{V(\hat r)}{c^2}.
\label{H_drop}
\end{equation}
Such a term follows from postulating the equivalence between mass and the energy \textit{operator} $mc^2\rightarrow mc^2+ \hat H_s$, which can be seen as a full \textit{quantum} implementation of the mass-energy equivalence. Both couplings, \eqref{H_PS} and \eqref{H_drop}, predict the same phase shift. In fact, all current observations are in agreement with effective theories, in which gravitational interactions of quantum particles do not include $\hat H_{vis}$, and which still correctly predict general relativistic effects in the classical limit and quantum effects in the Newtonian limit (see Appendix \ref{app_b} for such toy models and further discussion for their motivation). On the other hand, the experiment proposed in this work can distinguish the two couplings \eqref{H_PS} and \eqref{H_drop}, which cannot be done through (even arbitrary precise) measurements of the phase shift alone. The physical motivation of making such a distinction is to answer how the mass-energy equivalence extends to quantum mechanics and in particular how it applies to superpositions of different energy eigenstates. A summary of the concepts that are tested in different experimental setups is given in Table \ref{table:summary}.

The proposed quantum optics experiment may also allow testing some other non-standard theories of quantum fields \cite{tim}. Some of these alternative models predict a difference in the time evolution of entangled states in a curved background as compared to predictions of standard quantum filed theory in the same space-time (for flat space-time, such models reduce to the standard quantum field theory and are thus only distinguishable in a curved background). For example, the model proposed in Ref. \cite{tim} predicts a decorrelation of entangled photons, which can have a measurable effect on the expected visibility in our setup.

\renewcommand{\arraystretch}{1.2}
\captionsetup[table]{labelformat=default, labelsep=default, justification=centerlast, font=small}
\begin{table}[b]
\caption{Summary of the key interpretational aspects of our proposed experiments and of already performed experiments.
In order to test the overlap between quantum mechanics and general relativity, the observed effect cannot be compatible neither with classical mechanics nor with the Newtonian limit of gravity. An experiment in a given entry of the table and the indicated coupling are compatible with both theories that label the entry's row and column. The coupling of the gravitational potential $V(\hat r)$ to:  mass $m$, average energy  $\bra\hat H_s \ket$, and the energy operator $\hat H_s$ can be interpreted as testing the gravitational coupling due to:  Newton gravity, the semi-classical extension of the mass-energy equivalence, and the full quantum extension of the latter, respectively.
Note that some of the experiments have more than one possible interpretation, and thus fit in several different slots of the table. In contrast, our proposed clock interferometry experiment can only be placed in the slot of the table at the intersection of general relativity and quantum mechanics. Its successful realization would thus probe general relativity and quantum mechanics in a hitherto untested regime.\hspace*{\fill}}
\centering
\begin{tabular}{l c| lc| c| c|| c|}

												      & \multicolumn{1}{c}{} &              &                 \multicolumn{4}{ c }{$\;\;\;\;\;\;\;\;\;$ {\large{\bf gravity}}}   \Bspace       \\  \cline{5-7}
								 \TspaceBBig                  & \multicolumn{1}{c}{} &               &                     	                 &                              & non-metric,        &                \\
												      & \multicolumn{1}{c}{} &              &                    	                 &    non-metric,      & Newtonian  +    &  general relativity         \\
												      & \multicolumn{1}{c}{} &              &                                             &   Newtonian        &   semi-classical       &   (time  dilation)    \\    
												      & \multicolumn{1}{c}{} &              &   \Bspace                            &                              &  mass-energy equivalence       &       \\    \cline{3-7}

\multirow{3}{*}{\begin{sideways}{{\bf \large{mechanics \qquad}}}\end{sideways}} &  &  \multirow{2}{*}{classical} &  wave             &   \Bspace   \TspaceBig  \textit{well-tested}     &                                         &  \textit{Pound-Rebka \cite{ref:Pound1960}, Shapiro delay \cite{ref:Shapiro1971} }  \\ \cline{4-7}
                                                 							                            & \Bspace   \TspaceBig &         &    particle         & \textit{well-tested}
&    \textit{Pound-Rebka \cite{ref:Pound1960}}           &   \textit{Shapiro delay \cite{ref:Shapiro1971}}      \\ \cline {3-7} \vspace{-11pt} \\\cline{3-7}
     			    \TspaceBBig            					               &  &  \multirow{3}{*}{quantum}  & \multirow{3}{*}{ }   &   \textit{phase-shift of a }    &   \textit{ phase-shift of a   }    &   \textit{ quantum interference }\\      			                                                                   	&  &         &               &   \textit{matter wave \cite{ref:Colella1975,ref:Peters1999}}    &   \textit{single photon}          &   \textit{of clocks}    \\
                                                                                        &                       &        &                &                                                               & \textit{\textbf{[not yet tested]}}            & \textit{\textbf{[not yet tested]}}  \\ 
                                                                 \BspaceBig  &                       &        &                & \textit{(probes: $mV(\hat r)$)}                 & \textit{(probes: $\bra\hat H_s \ket \frac{V(\hat r)}{c^2}$)}            & \textit{(probes: $\hat H_s \frac{V(\hat r)}{c^2}$)}  \\  \cline {3-7}

\end{tabular}
\label{table:summary}
\end{table}

In order to qualify as a genuine test of quantum mechanics, the results of the experiment should be incompatible with a classical model of light. It is not sufficient that an arbitrary part of the experiment shows some quantum property (in fact, quantum mechanics is necessary to explain the atomic transitions of the clocks used in already performed tests of GR). What should be stressed is that the experiments performed so far are still compatible with models where classical, localized degrees of freedom ``keep track of time'' (see Appendix \ref{app_b}). Accessing the ``quantum domain'' means showing experimentally that the degree of freedom keeping track of time was in a genuine quantum superposition, with each part of the superposition evolving at a different rate due to GR. In our case an analogous interference with a modulation in the visibility could also be observed in the intensity of classical electromagnetic wave packets following the paths of the interferometer. However, within such a picture 
the phase at each point of the electromagnetic wave is a classical degree of freedom that keeps track of time.  An interference experiment in which a single photon source is utilized and single photons are detected is incompatible with this picture and thus would rule out an explanation of the experiment in terms of classical electromagnetic waves. The observation of quantum interference also rules out the model of photons as classical particles. As already noted, in the current proposal the signature of the general-relativistic time dilation is \textit{the lack} of interference. Implementation of an additional, controllable delay in the upper arm compensating the general relativistic time dilation would result in a recovery of the interference, which would prove the quantum nature of the photons affected by general-relativistic time dilation.

\subsection{Disproving local-realistic description of the experiment}
\begin{figure}[h!]
\vspace{0pt}
\begin{center}
\includegraphics[width=14.5cm]{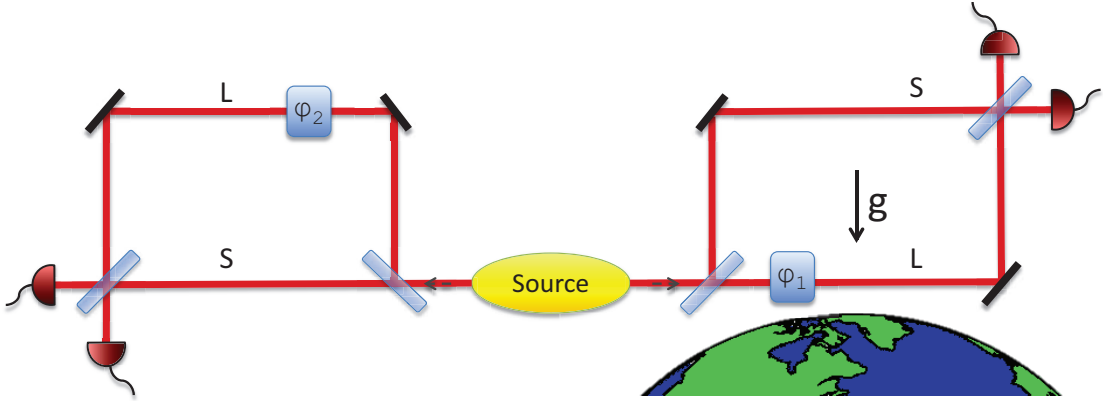}
\end{center}
\vspace{0pt}
\caption{A Franson-type interferometer modified to test a local realistic description of photons undergoing gravitational time dilation. It consists of a source of photon pairs and two Mach-Zehnder setups with arms of different optical lengths $S, L$.  On the right-hand side this difference is due to the general relativistic time dilation. This can be realized by utilizing a Mach-Zehnder setup as in Fig.\ \ref{machzehnder}, where the two paths of equal length are at different gravitational potentials, so that there will be a time dilation $\Delta \tau$ between them. The Mach Zehnder interferometer on the left-hand side is placed such that both arms feel the same gravitational potential and the difference in their lengths is $L-S = c \Delta \tau$. In each of the interferometers a controllable phase shift $\varphi_1$, $\varphi_2$ is induced in the optically longer arm. By varying those phases it is possible to violate Bell's inequalities on a subensamble of coincident detected photons.
\hspace*{\fill}}
\label{franson}
\end{figure}

The experimental scheme discussed so far can rule out models where light is described by classical electromagnetic waves or by classical particles evolving on a curved space-time\footnote{Note that classical models of light have already been extensively disproved experimentally, but only in regimes where no effect of GR is present.}. However, as any single photon experiment, it can still be described with a local realistic model. In order to rule out all local realistic models of photons in a curved background, a modified version of the experiment can be used. The idea is to use a Franson-type interferometer \cite{franson}, where the difference in the optical length between the two arms on one side of the interferometer is produced by the general relativistic time dilation (see Fig.\ \ref{franson}).
In this version of the experiment, pairs of time-correlated photons are emitted by a source and fed into two separated Mach-Zehnder interferometers. The right-hand side interferometer has arms of equal length and is positioned vertically with respect to the surface of the earth, thus the arrival time of the lower arm is delayed by an amount $\Delta \tau$, as given by Eq. \eqref{time_diff}. The left-hand side interferometer is horizontal, thus insensitive to gravity, but the two arms differ in their optical lengths by an amount $\Delta l= c\Delta \tau$; in this way both interferometers effectively consist of a shorter and a longer arm, labelled by $L$ and $S$ respectively, which pairwise have the same optical length. Two additional, controllable phases, $\varphi_1$ and $\varphi_2$, are added to the longer arms of the right and left interferometer, respectively.
The emission time of the two photons is correlated (to within their coherence time), but the specific emission time is unknown. For the coherence time of photons shorter than the time delay $\Delta \tau$, coincident photons will have the same time of flight.  As a result, the state of the post-selected fraction of coincident photons (just before the beam splitter) is entangled:
\begin{equation}
\label{timeentanglement}
|\psi \ket = \frac{1}{\sqrt{2}}\left(  e^{i(\varphi_1+\varphi_2)}|L\ket_1|L\ket_2 +|S\ket_1|S\ket_2\,\right).
\end{equation}
By varying the phases $\varphi_1$ and $\varphi_2$, it is possible to violate Bell's inequalities (even when taking into account the post-selection procedure \cite{marekonfranson}).
Thus, the scheme can provide a conclusive refutation of any model in which photons are described by local-realistic variables. At the same time, the violation of the inequalities is only possible if a photon travelling in the lower arm of the right-hand side interferometer is delayed by a time $\Delta \tau = \Delta l/c$. Since $\Delta l$ (the length difference in the left arm) can be controlled independently, it can be used to measure the time dilation on the right-hand side and verify the prediction of general relativity.

\section{Conclusion}
\label{sec:conclusion}
Probing the effect of gravitational time dilation on quantum interfering particles is a promising path towards fundamental tests of the overlap between quantum mechanics and general relativity. Here we discussed a quantum optics realization of this general idea, where a single photon travels in a superposition along the two paths of a Mach-Zehnder interferometer placed in the gravitational field. The gravitational time dilation between the two paths will cause the lower part of the superposition to be delayed as compared to  the upper one, leading to a loss of interference.
For a  total loss of interference a large-scale interferometer is required, which can be within reach in the near future. Already  within reach of present-day technology is, however, the measurement of the gravitationally induced phase shift for single photons.  Such an effect cannot be seen as a genuine test of gravitational time dilation, but it would probe a \textit{semi-classical} extension of the mass-energy equivalence in which the \textit{mean} energy couples to gravity, see Eq.\ \eqref{H_PS}. This feature was already observed in the Pound-Rebka experiment \cite{ref:Pound1960}, here, however, we propose to use single photons that travel in superposition to probe this effect in a regime where no classical description of light is possible.  In addition, observing the loss of quantum interference would corresponds to probing the full \textit{quantum} extension of the mass-energy equivalence in which the energy \textit{operator} couples to gravity, see Eq.\ \eqref{H_drop}. 
The observation of  general relativistic time dilation for  quantum interferring photons would represent an experimental test of the overlap between general relativity and quantum mechanics and would constitute a first step towards more advanced tests of the interplay between the two theories.

\begin{acknowledgments}
The authors thank S.\ Ramelov, M.\ Rupke and P.\ Walther for insightful discussions. The research was funded by the Austrian Science Fund (FWF) projects W1210 and SFB-FOQUS, the Foundational Questions Institute (FQXi), the European Commission Project Q-ESSENCE (No. 248095) and by the John Templeton Foundation.  M.Z., F.C. and I.P.\ are members of the FWF Doctoral Program CoQuS. M.Z.\ acknowledges hospitality at the Center for Interdisciplinary Research (ZiF), University of Bielefeld in March 2012.
\end{acknowledgments}

\appendix
\section{Evaluation of the expected photon number}
\label{app_a}

Consider the light source for the interferometer to be an ideal single photon source, which produces pulses propagating in the $+x$ (horizontal) direction each containing only one photon. In the local shell-frame on the earth's surface this can be represented by the quantum state \eqref{sig}. The use of plane-wave propagation is assumed justified by the paraxial approximation of a Gaussian spatial mode.

Detection of the horizontal output mode of the interferometer is assumed to be  broadband (i.e. a frequency band-width much greater than the source) and time integrated (detection time is much longer than the pulse width) and thus can be described by an operator
\begin{eqnarray}
a_o^{\dagger} a_o &=& \int {\frac{d\tau_{r'}}{2 \pi}} \int d\nu e^{i\frac{\nu}{c}(x_{r'} - c\tau_{r'})}  b^{\dagger}_{\nu} \int d\nu' e^{-i\frac{\nu'}{c}(x_{r'} - c\tau_{r'})}  b_{\nu'} =   \int d\nu  \; b^{\dagger}_{\nu}  b_{\nu},
\label{O}
\end{eqnarray}
where $b_{\nu'}$ are single wave-number boson annihilation operators for the output mode. We proceed by calculating the expectation value $\langle a_o^{\dagger}a_o \rangle$ against the initial state \eqref{sig} by finding the Heisenberg evolution of the detection operators. Quite generally the evolved single wave-number operators are of the form
\begin{eqnarray}
b_\nu &=& {\frac{1}{2}}a_\nu (e^{-i \nu\phi_{1}}-e^{-i \nu\phi_{2}})+  {\frac{1}{2}} v_{\nu} (e^{-i \nu\phi_{1}}+e^{-i \nu\phi_{2}}),
\end{eqnarray}
where $v_{\nu}$ are single wave-number boson annihilation operators from which the unoccupied input modes of the interferometer are constructed. Because they are initially in their vacuum state they will not contribute to the expectation value. The phases $\phi_{i}$, $i=1,2$ are acquired propagating along the corresponding paths $\gamma_i$ of the interferometer. We also require continuity at the mirror boundaries between the mode operator expressions along the different paths.

By symmetry, the contribution to the phases $\phi_i$ coming from the propagation along the radial part of the path is the same for both trajectories, as they are both evaluated over an equal time-interval and have the same lengths as measured by a distant observer. Because they are common they will be eliminated since only the phase difference $\Delta \phi := \phi_1 - \phi_2$ will contribute to the final expression. From  Eq.\ \eqref{sig} and the metric Eq.\ \eqref{M2} follows that the phases read $\phi_1 = \frac{1}{c}(l - c\tau_{r+h})$, $\phi_2 = \frac{1}{c}(l - c\tau_{r})$ and thus
\begin{equation}
\Delta \phi = \Delta \tau\, ,
\label{V}
\end{equation}
with $\Delta \tau$ given by the Eq.\ \eqref{time_diff}. Moreover,  the locally measured radial distance $h$ between the paths is  found via
\begin{eqnarray}
h_r = \int^{r+h}_r {\frac{dr'}{\sqrt{1-\frac{2GM}{c^2r'}}}}\;.
\label{rob}
\end{eqnarray}
Evaluating the photon number expectation value for the state $|1 \rangle_f =\int d\nu f(\nu) e^{i\frac{\nu}{c}(x_{r} - c \tau_{r})} a^{\dagger}_{\nu} |0 \rangle$, Eq.\ \eqref{sig},  yields
\begin{eqnarray}
\langle a_o^{\dagger} a_o \rangle &=& \langle 1|_f  \int d\nu  \; b^{\dagger}_{\nu}   b_{\nu} \; |1 \rangle_f  =  \int d\nu |f(\nu)|^2 {\frac{1}{4}} |1 - e^{i \nu \Delta \tau}|^2 \;,
\label{bcb}
\end{eqnarray}
which is the same result as that of Eq.\eqref{photon_probab} for a single photon normalized wave packet.

\section{Toy models}
\label{app_b}
Both experimental proposals  discussed in this work are formulated within the framework of quantum mechanics in curved space-time. No effects specific to this theory have been experimentally verified so far - bridging this gap remains the principal motivation behind our work.  Quite generally, new physics is expected only at the scale where gravity itself could no longer be described within a classical theory. However, the tension between quantum mechanics and general relativity is of conceptual nature. Both theories stress that only operationally well defined notions may have physical meaning and this concerns also the notion of time (or proper time in general relativity).  However, in contrast to general relativity, in quantum mechanics any degree of freedom of a physical system can be in a superposition and thus becomes undefined (beyond the classical probabilistic uncertainty). More generally -  the theory allows for physical states that cannot be described within any local realistic model. If this applies to the degrees of freedom  on which our operational treatment of time relies -  the latter becomes classically undefined. This can be the case even when space-time itself can still be described classically, like in the proposals discussed in this paper. 
One could, however, take an opposite view and assume that whenever space-time itself is classical, the time for any system, that constitutes an operationally defined clock, should admit a classical description as well. The tension between these two views motivates the investigation of theoretical frameworks alternative to quantum mechanics in curved space time. Here we sketch an explicit example of such an alternative toy theory, which can be tested by the experiment proposed in this work.

Relevant for the present problem is how the physical degrees of freedom evolve on a curved background. In the standard approach such evolution results in entanglement between the spatial mode of the wavefunction and  other degrees of freedom. There is no well-defined time that such degrees of freedom experience and even a Bell-type experiment can be designed in which any local realistic model of time can be refuted \cite{we_in_preparation}.
This entanglement results from the coupling $\hat H_{vis}= \hat H_s\frac{V(\hat r)}{c^2}$, see  Eq.\ \eqref{H_drop}, and is the reason for the drop in the interferometric visibility (for both massive and massless cases of the experimental proposal discussed in this work). 
All so far observed gravitational effects can, however, be explained with one of two possible effective forms of such an interaction, which reproduce only specific features of \eqref{H_drop} and correspond to different physical effects.

The effective coupling $\hat H_{phase}=\bra \hat H_s\ket \frac{V(\hat r)}{c^2}$, see  Eq.\ \eqref{H_PS}, reproduces correctly the gravitational phase shift effect of the standard theory, but not the time dilation.
Applying the operator \eqref{H_PS} to the state of the clock degree of freedom $|\tau\ket$ in a spatial superposition of two locations $r_1$ and $r_2$,
$|\Psi\ket=\frac{1}{\sqrt{2}}\left(|r_1\ket + |r_2\ket \right) |\tau\ket$, we get
$$\hat H_{phase}|\Psi\ket = \frac{\bra \hat H_s\ket}{\sqrt{2}}\left(\frac{V(r_1)}{c^2}|r_1\ket + \frac{V(r_2)}{c^2}|r_2\ket \right) |\tau\ket,$$
where $\bra \hat H_s\ket = \bra \tau| \hat H_s |\tau\ket$. The full evolution in such a toy model is given by the Hamiltonian $\hat H_s + \hat H_{ps}$, which yields
\begin{equation}
\label{ps_evolution}
|\Psi(t)\ket=e^{-\frac{i}{\hbar}\left(\hat H_s + \hat H_{phase}\right)t}|\Psi\ket = \frac{1}{\sqrt{2}}\left[e^{-\frac{i}{\hbar}\left(\bra \hat H_s\ket \frac{V(r_1)}{c^2}\right)t}|r_1\ket + e^{-\frac{i}{\hbar}\left(\bra \hat H_s\ket \frac{V(r_2)}{c^2}\right)t}|r_2\ket \right] |\tau(t)\ket,
\end{equation}
where $|\tau(t)\ket=e^{-\frac{i}{\hbar}\hat H_s t}|\tau\ket$
and thus the evolution of the clock degree of freedom $|\tau\ket$ does not depend on the position $r$. Hence, the coupling \eqref{H_PS} predicts no general relativistic time dilation and no drop in the interferometric visibility -  the clock degree of the freedom remains factorized from the spatial modes. However, each mode acquires a phase proportional to the gravitational potential and an effective mass defined by $\bra \hat H_s\ket$, hence 
this effective coupling reproduces the relative phase shift measured in interference experiments.  For a particle of mass $m$, $\hat H_s = mc^2 + \hat H_{int}$ (to first order in $1/c^2$), where $\hat H_{int}$ is the Hamiltonian of the internal degrees of freedom
and thus one obtains $m + \frac{\bra \hat H_{int}\ket}{c^2}$ for the effective mass. The first term is simply the Newtonian mass, while the second is a relativistic correction. For a single photon mode with a frequency $\omega$, $\hat H_{s}=\hbar\omega a^{\dag}a$ and therefore the whole contribution to the phase shift comes from $\frac{\bra \hat H_{s}\ket}{c^2}= \frac{\hbar\omega}{c^2}$. Thus, any measurement of the gravitational phase shift for photons would represent a signature of a non-Newtonian effective mass. However, no measurement of the phase shift (in the massive or massless case) could represent a measurement of the time dilation, since the phase shifts are explainable by the coupling \eqref{H_PS} which does not cause clocks at different potentials to tick at different rates.

A different effective coupling can explain all classical general relativistic effects observed so far. It includes an effective gravitational potential $\bra V(\hat r)\ket$ - gravitational potential smeared over the support of the wavefunction of a single physical system. Such a coupling reads
\begin{equation}
\hat H_{loc} = \hat H_s\frac{ \bra V(\hat r)\ket}{c^2}
\label{HeffGR}
\end{equation}
and it accounts for gravitational experiments in which the relevant degrees of freedom are sufficiently well localized. These include not only classical tests of general relativity \cite{ref:Pound1960, ref:Hafele1972, ref:Shapiro1964, ref:Shapiro1971}, but also experiments measuring the time dilation between two  \textit{localized} atomic clocks, each at a different gravitational potential. 

More generally, one can construct a toy model by combining the above effective couplings $\hat H_{phase}$ and $\hat H_{loc}$, for example
\begin{equation}
\hat H^{eff}_i = \hat H_s\left(1+\frac{\bra V(\hat r)\ket}{c^2}\right)+\frac{\Delta N_i}{\bra \hat N_i\ket} \left(\bra \hat H_s \ket\frac{V(\hat r)}{c^2} - \hat H_s \frac{\bra V(\hat r)\ket}{c^2}  \right),
\label{Heff}
\end{equation}
which governs the evolution of the $i^{th}$ mode of a quantum state (one can associate different modes to e.g., different paths of a Mach-Zehnder interferometer).  $\hat N_i$ is the number operator in mode $i$ and $\Delta N_i$ is its standard deviation. The parameter $\frac{\Delta N_i}{\bra \hat N_i\ket}$ quantifies how well  the quantum state is localized. It vanishes for Fock states (e.g. for a pair of atoms or photons, each in one, localized mode), and in the limit of large coherent states (for a coherent state  $|\alpha\ket$ in mode $i$ we have $\frac{\Delta N_i}{\bra \hat N_i\ket}=\frac{1}{|\alpha|}\rightarrow 0$ for $\alpha\rightarrow\infty$), which for photons corresponds to  classical light.  In both cases the Hamiltonian \eqref{Heff} reduces to  $\hat H_s\left(1+\frac{\bra V(\hat r)\ket}{c^2}\right)$. In the other limit, when the parameter $\frac{\Delta N_i}{\bra \hat N_i\ket}=1$, which is the case for a single particle in a superposition of two modes, the effective Hamiltonian reduces to $\hat H_s+\bra \hat H_s\ket\frac{ V(\hat r)}{c^2}$.  The toy model \eqref{Heff} predicts no drop in the interferometric visibility for a particle in a spatial superposition (since in the relevant limit the energy \textit{operator} $\hat H_s$ does not couple to the potential) but is still consistent with the experiments carried out so far. (Moreover, it can be generalized beyond the above weak energy limit.) 

The difference between the standard extension of quantum mechanics to curved space-time and the toy model \eqref{Heff}  can only be tested with a quantum system from which the time can be read out and which is put in a coherent spatial superposition at different gravitational potentials.  Even though this model is artificial (e.g., it shares the difficulties of all quantum nonlinear models) it highlights the conceptual difference between gravitational phase shift experiments and measurements of the visibility loss. (While the former only probe the semi-classical coupling of energy to the gravitational potential, the latter directly test the full quantum form of such a coupling.)  Most importantly, the toy model emphasizes the necessity of probing quantum mechanics in curved space-time:  the results of current experiments cannot necessarily be extrapolated to this regime.

\end{document}